\documentclass[12pt, preprint]{aastex}
\begin{document}

\title{X-ray Pulsars in the Small Magellanic Cloud}
\author{W.A.Majid}

\affil{Jet Propulsion Laboratory, California Institute of Technology,
Pasadena, CA 91109}

\author{R.C.Lamb}
\affil{Space Radiation Laboratory, California Institute of Technology,
Pasadena, CA 91125}

\author{D.J.Macomb}
\affil{Physics Department, Boise State University, Boise, ID 83725-1570}

\begin{abstract}

XMM-Newton archival data for the Small Magellanic Cloud have been
examined for the presence of previously undetected X-ray pulsars.  One
such pulsar, with a period of 202 s, is detected.  Its position is
consistent with an early B star in the SMC and we identify it as a
high mass X-ray binary (HMXB).  In the course of this study we
determined the pulse period of the possible AXP CXOU J010043.1-721134
to be 8.0 s, correcting an earlier report (Lamb et al 2002b) of a 5.4
s period for this object.  Pulse profiles and spectra for each of
these objects are presented as well as for a recently discovered
(Haberl \& Pietsch 2004) 263 s X-ray pulsar.  Properties of an
ensemble of 24 optically identified HMXB pulsars from the SMC are
investigated.  The locations of the pulsars and an additional 22 X-ray
pulsars not yet identified as having high mass companions are located
predominately in the young (ages $\le 3 \times 10^{7}$ years) star
forming regions of the SMC as expected on the basis of binary
evolution models. We find no significant difference between the
distribution of spin periods for the HMXB pulsars of the SMC compared
with that of the Milky Way.  For those HMXB pulsars which have Be
companions we note an inverse correlation between spin period and
maximum X-ray flux density.  (This anti-correlation has been
previously noted for all X-ray binary pulsars by Stella, White \&
Rosner 1986).  The anti-correlation for the Be binaries may be a
reflection of the fact that the spin periods and orbital periods of Be
HMXBs are correlated.  We note a similar correlation between X-ray
luminosity and spin period for the Be HMXB pulsars of the Milky Way
and speculate that exploitation of the correlation could serve as a
distance indicator.

\end{abstract}

\section{Introduction}

The advantages to studying the compact object content of the
Magellanic Clouds are well known: a relatively small angular size, at
a reasonably close and known distance, located at a high galactic
latitude with little obscuration by the interstellar medium of the
Galaxy.  In addition, there is evidence for recent star formation in
the SMC over the last few 10's of millions of years which creates an
environment in which X-ray pulsars are expected to be plentiful
(Haberl \& Sasaki 2000, Maragoudaki et al. 2001).  This expectation is
borne out by recent discoveries of more than a score of previously
unknown X-ray pulsars in the SMC based on observations from the ASCA,
ROSAT, RXTE, and Chandra X-ray missions.  Recent reviews of these
discoveries are presented in Yokogawa et al. (2003) and Haberl
\& Pietsch (2004).

Our analysis of recently available archival data from the {\it X-ray
Multi-Mirror Mission--Newton} satellite reveals evidence for another
X-ray pulsar from the direction of the SMC, having a period of 202 s.
It is sufficiently close to a high mass companion that the association
is likely.  This discovery continues our program of routinely
evaluating the pulsar content of archival data from the SMC.  This
monitoring has revealed eight previously undiscovered X-ray pulsars
(Macomb et al. 1999, Finger et al. 2001, Lamb et al. 2002a, Lamb et
al. 2002b, Macomb et al 2003).  Analysis of this same archival dataset
shows that the possible Anomalous X-ray Pulsar (AXP), CXOU
J010043.1-721134, (Lamb et al. 2002b,2003) has a pulse period of 8.02
s, correcting an initial report of a 5.4 s period for this object.
Characteristics which support its likely identity as an AXP are
presented.  In the section 2 we present the X-ray observations and
spectral analysis relating to the 202 s pulsar and the possible AXP as
well as detailed pulse phase spectroscopy of the recently discovered
263 s pulsar (Haberl \& Pietsch 2004).  In section 3 we present
evidence relating to the optical identification of the 202 and 263 s
pulsars, followed by a discussion, in section 4, of these pulsars as
well as the possible AXP,

X-ray binaries are conventionally divided into two distinct classes:
high mass (companion stars with masses $\ge$ 8 solar masses) and low
mass (companion masses $\le$ 1 solar mass) with relatively few X-ray
binaries having companions with masses in the intermediate range.  The
division into the two classes makes sense, since their evolution is
quite different.  For the high mass X-ray binaries (HMXBs), the
maximum lifetime for their existence is set by the nuclear burning
time of an 8 solar mass star, which is $\sim 3\times 10^7 $ yr
(Verbunt \& van den Heuvel 1995).  Whereas low-mass X-ray binaries
have lifetimes of $10^{8-9}$ yr or more.  This characteristic of the
HMXBs makes them a useful measure of recent star forming activity in
galaxies (see for example: Grimm et al. 2003).

The HMXB population of the Small Magellanic Cloud has recently been
catalogued by Liu et al. (2000).  Since then two papers have reported
on the X-ray population of the SMC based on specific X-ray satellite
data.  Yokogawa et al. (2003) summarizes source populations bases on
ASCA data; Sasaki et al. (2003) on the basis of {\it X-ray
Multi-Mirror Mission--Newton} satellite data.  In section 5 we present
an overview of the HMXBs of the SMC.  We note (as other authors have
noted) that virtually all of are located in young (ages $\le 3\times
10^7$ years) star forming regions as expected on the basis of binary
evolution models (Verbunt \& van den Heuvel 1995).  We note a
significant inverse correlation between period and maximum X-ray flux
density for those HMXBs with Be companions.  (This anti-correlation
has been previously noted for all X-ray binaries by Stella, White \&
Rosner 1986.)  The anti-correlation may be a reflection of the fact that
for such objects spin period is correlated with orbital period (Corbet
1986) and longer period Be HMXBs have lower maximum accretion rates.
We note a similar correlation between X-ray luminosity and spin period
for the Be HMXB pulsars of the Milky Way.  We find no significant
difference between the distribution of periods from the HMXBs of the
SMC compared with that of the Milky Way, except for a slight deficit
for SMC pulsars with periods longer than $\sim$200s.  This may be due
to their generally smaller luminosity compared to the luminosity of
SMC HMXB pulsars with faster spin.

\section{X-ray Observations}

Table 1 gives a list of the XMM-Newton archival data which were
surveyed for this paper.  Data from the European Photon Imaging
Cameras EPIC PN (Str\"uder et al. 2001), and MOS1 and MOS2 (Turner at
al. 2001) were utilized.  For the first stage of the analysis, we
utilized the data products of the Pipeline Processing Subsystem (PPS
files) to search for pulsars.  Point-like sources were identified from
image files and circular selection regions around each source were
chosen by eye.  Radii of selection circles varied from 15 to 32.5 arc
seconds depending on source strength and position in
the field of view.  Sources with fewer than $\sim$ 500 photons were
not examined.  With this cut-off a total of 63 sources were selected for
timing analysis.  

\begin{deluxetable}{lcccccc}

\tablecolumns{7}

\tablecaption{Observations of the SMC by XMM-Newton searched for X-ray pulsars}

\tablewidth{0pt}

\tablehead{\colhead{Obsid ID } & \colhead{RA (J2000)} & \colhead{DEC
(J2000)} & \colhead{Comments} & \colhead{Instrument} & \colhead{Time
res.} &\colhead{Duration(ks)}}

\startdata
0018540101 & 00 59 26.8 &  -72 09 55 & Possible AXP & M1,M2 &  2.6s &27.3 \\
	&	&	&	     &  PN & 73.4ms & 19.1 \\
0084200101 & 00 56 41.7 &  -72 20 24 &  & M1,M2 & 2.6s & 21.4  \\
	&	&	&	     &  PN & 73.4ms & 19.1 \\	       
0084200801 & 00 54 31.7 &  -73 40 56 &  & M1,M2 & 2.6s & 22.0 \\
        &       &       &            &  PN & 73.4ms &18.5 \\
0110000101 & 00 49 07.0 &  -73 14 06 &  263 s Pulsar & M1,M2 & 2.6s & 26.9 \\
	&	&	&	     &  PN  & 200ms  & 23.0 \\
0110000201 & 00 59 26.0 &  -72 10 11 &  202s Pulsar & M1,M2 & 2.6s & 19.7 \\
0110000301 & 01 04 52.0 &  -72 23 10 & & M1,M2 & 2.6s & 32.6 \\
	&	&	&	     &  PN  & 200ms & 31.0 \\
0135720601 & 01 03 50.0 &  -72 01 55 & & M1,M2 & 2.6s & 32.9 \\
        &       &       &            &  PN  & 6ms & 31.0 \\
0135720801 & 01 04 00.0 &  -72 00 16 & & M1,M2 & 2.6s & 32.2 \\
        &       &       &            &  PN  & 6ms & 31.3 \\
0135720901 & 01 04 01.7 &  -72 01 51 & & M1 & 1.5ms & 13.0 \\
        &       &       &            &  M2  & 0.9s & 13.0 \\
        &       &       &            &  PN  & 73.4ms & 13.9 \\
0135721001 & 01 04 01.7 &  -72 01 51 & & M1 & 1.5ms & 13.1 \\
        &       &       &            &  M2  & 0.9s & 14.9 \\
        &       &       &            &  PN  & 73.4ms & 19.4 \\
0135721301 & 01 03 56.4 &  -72 00 28 & & M1,M2 & 0.9s & 28.7  \\
        &       &       &            &  PN  & 6ms & 10.9 \\
0011450101 & 01 17 05.1 &  -73 26 35 &  SMC X-1 region & M1,M2 & 2.6s & 59.0 \\
        &       &       &            &  PN  & 73.4ms & 54.0 \\
0011450201 & 01 17 05.1 &  -73 26 35 &  SMC X-1 region & M1,M2 & 71.5ms & 40.7 \\
        &       &       &            &  PN  & 1.5ms & 40.2 \\

\enddata
\tablecomments{M1 and M2 refer to EPIC detector MOS1 and MOS2.}
\end{deluxetable}

The selection circles for each source provide a spatial filter to make
photon event lists from the PPS event files. The times of arrival for
these photons were corrected to the barycenter of the solar system
using the SAS software.  Photon arrival time information was then used
in a search for significant periodicities at frequencies less than the
Nyquist limit of each dataset. This limit depends on the data mode of
each instrument.  Generally for PN data, the Nyquist limit was 6.8 Hz,
for the MOS data, the limit was generally 0.19 Hz. The time resolution
for each of the EPIC instruments for each of the observations is given
in Table 1.  Initial Fourier transforms were performed separately on
the two data types, PN and MOS.  In the first pass through the data
there was no screening for possible background flaring events.  For
the final analysis of the three pulsars described in this paper,
contamination from possible background flaring events was searched
for, but not found.

In the first pass through the data no significant frequency less the
Nyquist limit was detected.  To increase the sensitivity to a possible
signal, data from all three detectors (MOS1, MOS2, and PN if
available) were then merged for further analysis.  With this combined
data, three different energy selections were made: all energies,
energies above 1 keV, and energies above 2 keV. With this procedure
two significant periodicities were found, 202 s and 263 s.  (During
the course of our investigation, the 263 s periodicity was reported by
Haberl \& Pietsch (2004).)  In addition to these periodicities, seven
previously known pulsars were also detected: pulsars with periods of
172, 323, and 755~s in the field of observation 0110000101, pulsars
with periods of 152, 280, and 452 s in the field of 0110000201, and
the possible AXP, CXOU J010043.1-721134, with a pulse period of 8.02 s
rather than the previously reported 5.4 s (Lamb et al. 2002b).  The
details of for both the 202 s and 263 s pulsars as well as for the
possible AXP are reported in this paper.

The event selection for our spectroscopic studies of these three sources
utilized the XMM-Newton Science Analysis System (SAS) (version 5.3)
software from the XMM-Newton Science Operations Center (SOC). The data
were filtered to remove ``bad'' pixels and events falling at the edges
of the CCDs.  We retained only standard XMM-Newton event grades
(patterns 0-12 for MOS1 and MOS2, and patterns 0-4 for the PN
instrument).

A spectrum was extracted, using the {\it evselect} program in SAS,
grouping photons in bins of 15 eV
for the MOS instruments and 5 eV for the PN instrument.  An energy cut
was applied to include only events in the energy range 0.25-10.0 keV.
For estimating the background, we chose several source free regions in
the same CCD as our target source.  The results of the spectral
analysis were not sensitive to the choice of the background region
chosen.  All spectral fits employed the XSPEC v11.2  software fitting
package.

\subsection{1XMMU J005921.0-722317. 202 s Pulsar }

This pulsar was discovered in an analysis of observation 0110000201
for which only MOS data were available in the public archive.  This
observation is rich in the proportion of sources which have been
established as binary X-ray pulsars.  Of the seven sources in this
field which have more than 500 photons, 4 have been previously
reported to be X-ray pulsars.  Our timing analysis detected 3 of these
previously known pulsars, only the possible AXP, CXOU J010043.1-721134
(Lamb et al. 2002b) was not detected.  (However clear pulsations from
this latter source were detected from another observation, 0018540101,
and this detection is discussed in a subsection below.)  The position
of the 202 s pulsar (1XMM J005921.0-722317) taken from (Sasaki,
Pietsch \& Haberl 2003) is: R.A. = 00$^h$ 59$^m$ 21.05$^s$, Dec. =
-72$^\circ$ 23$'$ 17.1$''$ (J2000).

For 1XMMU J005921.0-722317, no significant Fourier power was detected
in either the uncut data, nor the data selected to have photon
energies greater than or equal to 1.0 keV.  However, for photons with
energies greater than 2.0 keV, highly significant normalized power of
26.6 (chance probability, after consideration of the frequency and
energy search ranges of less than $6\times10^{-7}$).  Optimization of
the region and energy selected gave a final power of 30 at 0.004954
(12) Hz, with a final energy cut of 1.9 keV and a circle of selection
of 27.5 arc seconds.  There is small, but significant power (4) at the
second harmonic.  The barycentered frequency of the fundamental
corresponds to a pulse period of 201.9 $\pm$0.5 s.  Figure 1 (a) shows
a portion of the resulting Fourier transform for this source.  Figure
1 (b) shows the pulse profile at this frequency for those photons with
energies greater than 1.9 keV.  The pulsed fraction with this energy
selection is: 50$\pm$8\%.  The epoch of the observation is 51834.633
MJD.  In Figure 1 (c) the pulse profile for photons with energies
below 1.9 keV is shown, illustrating the lack of any significant
pulsation for this energy selection.

This source also present in observation 008420001, 528.9 days later
than observation 0110000201.  In this observation the source flux is
70\% of that of the earlier observation.  A Fourier transform of the
source photons with a similar energy selection as that done for
0110000201 reveals a power of 15 at a frequency of 0.004968 (12) Hz
(201.2$\pm$0.5 s pulse period).  Although this is not a highly
significant detection, we regard it as supporting evidence for the 202
s pulsation.  The difference in frequency between the two observations
is 14$\pm$17 $\mu$Hz which is not significant.  An upper limit on a
frequency derivative over the interval between the two observation is
$3 \times 10^{-13}$ Hz/s.

\begin{figure}
\figurenum{1} \epsscale{1.0} \plotone{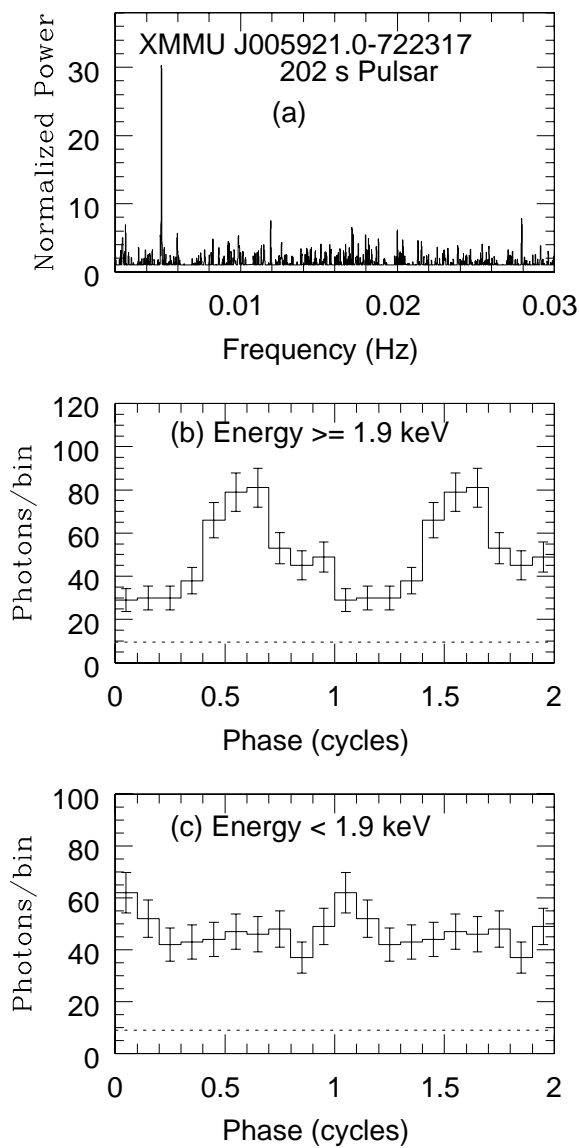}

\caption{(a) A portion of the Fourier transform for 1XMMU
J005921.0-722317 using combined MOS data for photon energies greater
than or equal to 1.9 keV.  The peak frequency is 0.004954 (12) Hz
corresponding to a period of 201.9 s. (b) Pulse profile of 1XMMU
J005921.0-722317 for photon energies above 1.9 keV.  The dashed line
indicates the non-source background level.  (c) Pulse profile for
photon energies below 1.9 keV.}

\end{figure}

The spectral data from the two MOS instruments were fitted
simultaneously.  The spectrum is best fit with a simple powerlaw
model, modified by neutral absorption.  Each model parameter is
allowed to vary freely, but was constrained to maintain the same value
for both spectra.  An overall normalization parameter for each data
set is allowed to vary to account for the slightly different count
rates in each detector.  With this model we obtain an acceptable
$\chi^{2}/\nu$ = 0.996 for 20 degrees of freedom.  Best fit parameter
values for this model are listed in Table 2.  

We also attempted to fit a model consisting of powerlaw and blackbody
components, modified by neutral absorption.  We obtained a somewhat
higher, yet still adequate, $\chi^{2}/\nu$ = 1.10 for 18 degrees of
freedom.  However, the blackbody temperature obtained is unreasonably
high and is not constrained by this fit ($3 \pm 60 \, keV$). We note
that a simple blackbody model results in a very poor fit
($\chi^{2}/\nu$ = 3.6 for 18 dof).

\subsection{Possible AXP, CXOU J010043.1-721134}

This source was first reported as a 5.4 s (frequency 0.184 Hz) pulsar
by Lamb et al. (2002b) from an analysis of {\it Chandra} ACIS-I data.
In the Fourier transform of the {\it Chandra} data two power peaks
appeared, one at 0.125 Hz just below the ACIS-I Nyquist frequency, and
the other at 0.184 Hz.  The two peaks are aliases of each other. Our
efforts to remove this frequency ambiguity used ROSAT observations and
concluded (erroneously) that the correct frequency was 0.184 Hz.  Our
present analysis of XMM data using PN data from observation 0018540101
decisively settles the question of the pulse period.  The PN data has
a Nyquist frequency of 6.8 Hz, far above the region of interest.

A selection circle 30 arc seconds in radius around the source region
produced 4048 photons.  The Fourier transform of these times showed a
strong, normalized power 28, peak at a frequency of 0.124699(5) Hz.  A
selection of photons with energies less than 6.0 keV increased this
power to more than 30.  The FFT of these energy selected photons is
shown in Figure 2 (a).  In addition to the peak at the fundamental
there is a clear signal at the second harmonic as well as a much
weaker signal (power 11) at the third harmonic.  The pulse profile for
these photons at this frequency is shown in Figure 2 (b).  The pulsed
fraction is 33$\pm$5\%.

The epoch for this observation is 52234.015 MJD.  The initial {\it
Chandra} observation (Lamb et al. 2002b) occurred 189.934 days earlier
and the observed frequency for that epoch was 0.124706(1) Hz.  By
combining the two observations we derive a frequency time derivative
of -4$\pm3$ $\times$10$^{-13}$ Hz/s.  This corresponds to a
characteristic timescale, $P/\dot P$, of $\sim$ 10 kyr, consistent with
the similar timescales observed for other AXPs (See Mereghetti et al. 2002
for a review).

\begin{figure}
\figurenum{2} \epsscale{1.0} \plotone{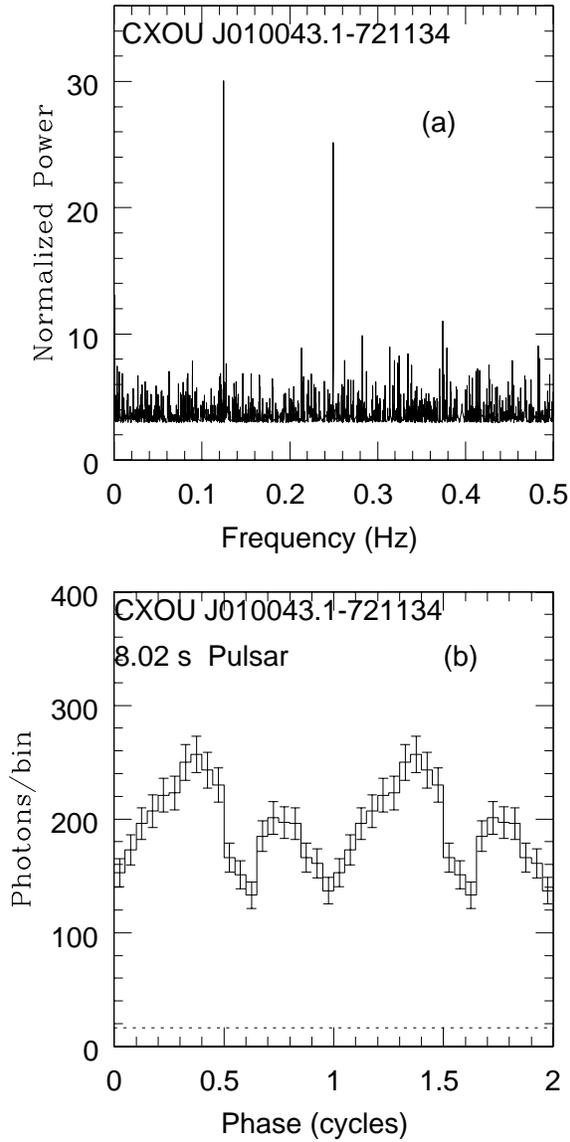}

\caption{(a)A portion of the Fourier transform for CXOU
J010043.1-721134 (1XMMU J010043.1-721134).  The peak
at 0.1248 Hz corresponds to the fundamental frequency of the possible
AXP.  Also second and third harmonics are visible as well.  the (b)
Pulse Profile of CXOU J010043.1-721134. The dashed line indicates the
non-source background level.}

\end{figure}

The spectrum for this source was fitted to the PN data of observation
0018540101. The spectrum of the source is well fit with a simple
powerlaw or a combination of powerlaw and blackbody components.
However, the fit using the combination of powerlaw and blackbody did
not constrain the normalization of either component, with the size of
the errors on each normalization comparable to the value of the
normalizations.  Figure 3 shows the best fit to the data using the
simple powerlaw model, modified by neutral absorption.  The best fit
parameters are given the Table 2.

\begin{figure}
\figurenum{3}

\includegraphics*[angle=270]{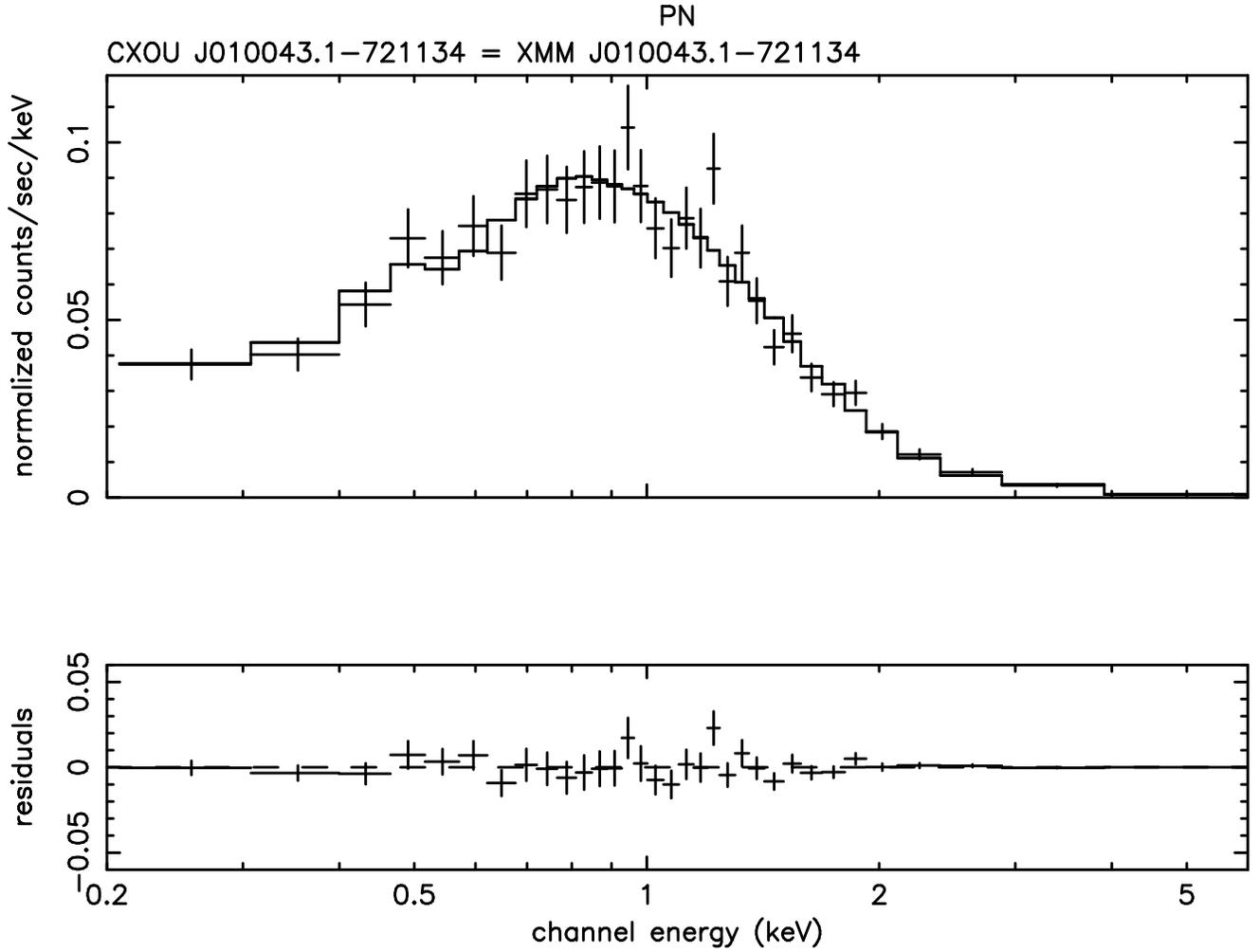}
\caption{Best fit spectrum for CXOU J010043.1-721134 for a model with
a powerlaw modified by neutral absorption.  The parameters for
this model are given in Table 2.}

\end{figure}

\subsection{1XMMU J004723.7-731226. 263 s Pulsar }

This pulsar was found in the analysis of observation 0110000101, a
field rich in X-ray pulsars.  Recently Haberl \& Pietsch (2004)
reported detection of 263 s pulsations from this same source.
Thirteen sources in observation 0110000101 meet the criterion of a
minimum of 500 photons, including three previously detected as
pulsars.  Our analysis detected pulsations from all three.  The
position of the 263 s pulsar taken from (Haberl \& Pietsch 2004a) is:
R.A. = 00$^h$ 47$^m$ 23.7$^s$, Dec. = -73$^\circ$ 12$'$ 26.9$''$
(J2000).

The initial selection circle for J004723.7-731226 was 25 arc seconds.
Significant power was separately detected at a fundamental frequency
of 0.0038 Hz for both the MOS1 and MOS2 combined data (normalized power
28) and PN data (normalized power of 29).  These Fourier power values
are consistent with the near equality of the number of selected
photons for the two different types of detector.  In the Fourier
analysis of the combined dataset (PN plus MOS) a normalized power of 56
was obtained with evidence for significant power for the next three
higher harmonics.

In order to maximize our sensitivity to the pulsed signal two
optimization procedures were further employed.  By using the strength
of the Fourier power at the fundamental plus the next four harmonics
as a criterion, a search on position and radius of a selection circle
was performed.  A search was also made to establish the optimum energy
selection.  After these two optimizations, a normalized Fourier power
(fundamental) of 81 resulted, with a power of 150 for the sum of the
first five harmonics.  The optimum selection circle for the PN data
was 27.5 arc seconds, for the MOS data 32.5 arc seconds.  The
optimum energy selection required the photon energy to be between
0.1 keV and 5.0 keV.

Figure 4 shows the resulting optimized Fourier transform for
J004723.7-731226.  The fundamental, barycentered frequency is
0.003803~(3) Hz, with significant power clearly visible at the second,
third and fourth harmonics.  This frequency corresponds to the pulse
period of 263.0$\pm$0.2 s.  The epoch of the observation is
51832.684 MJD.  Figure 5(a) shows the pulse profile corresponding
to the data of Figure 4.  The pulsed fraction is 47$\pm$4\%.

\begin{figure}
\figurenum{4} \epsscale{1.0} \plotone{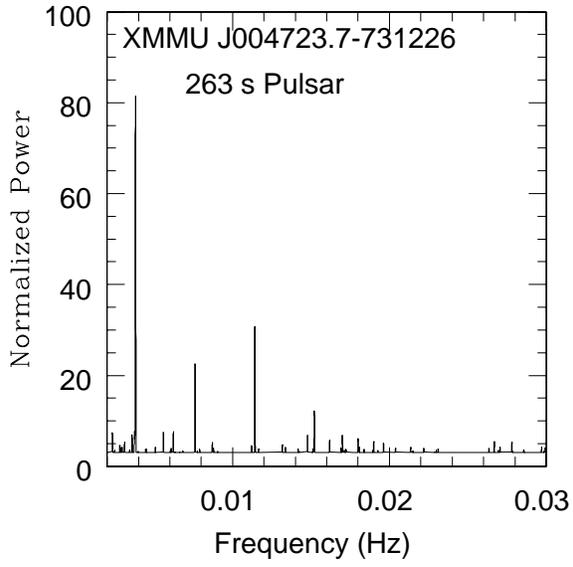}
\caption{A portion of the FFT for 1XMMU J004723.7-731226 using combined
PN and MOS data.
In addition to power at the fundamental frequency of 0.003803(3) Hz, there
are clear power peaks at the second, third, and fourth harmonics.}

\end{figure}

\subsubsection{Pulse Profile for Different Energy Bins}

Using the best period and combining the data from the three
instruments, we calculated a phase for each event passing our event
selection criterion.  The pulse profile can be characterized as a
broad peak, with one or possibly two dips within this overlying
envelope.  To search for any energy dependent features in the folded
pulse profile, four contiguous energy intervals were chosen: I:
0.3-1.0 keV, II: 1.0-2.5 keV, III: 2.5-5.0 keV, IV: 5.0-10.0 keV (see
Figure 5).  We can identify the first dip in the broader pulse profile
in all but the last energy interval (IV).  We note that there is a
marginal feature near the second dip in the 0.3-1.0 keV interval,
which could simply be due to statistical fluctuations.

\begin{figure}

\figurenum{5}
 \epsscale{1.0} \plotone{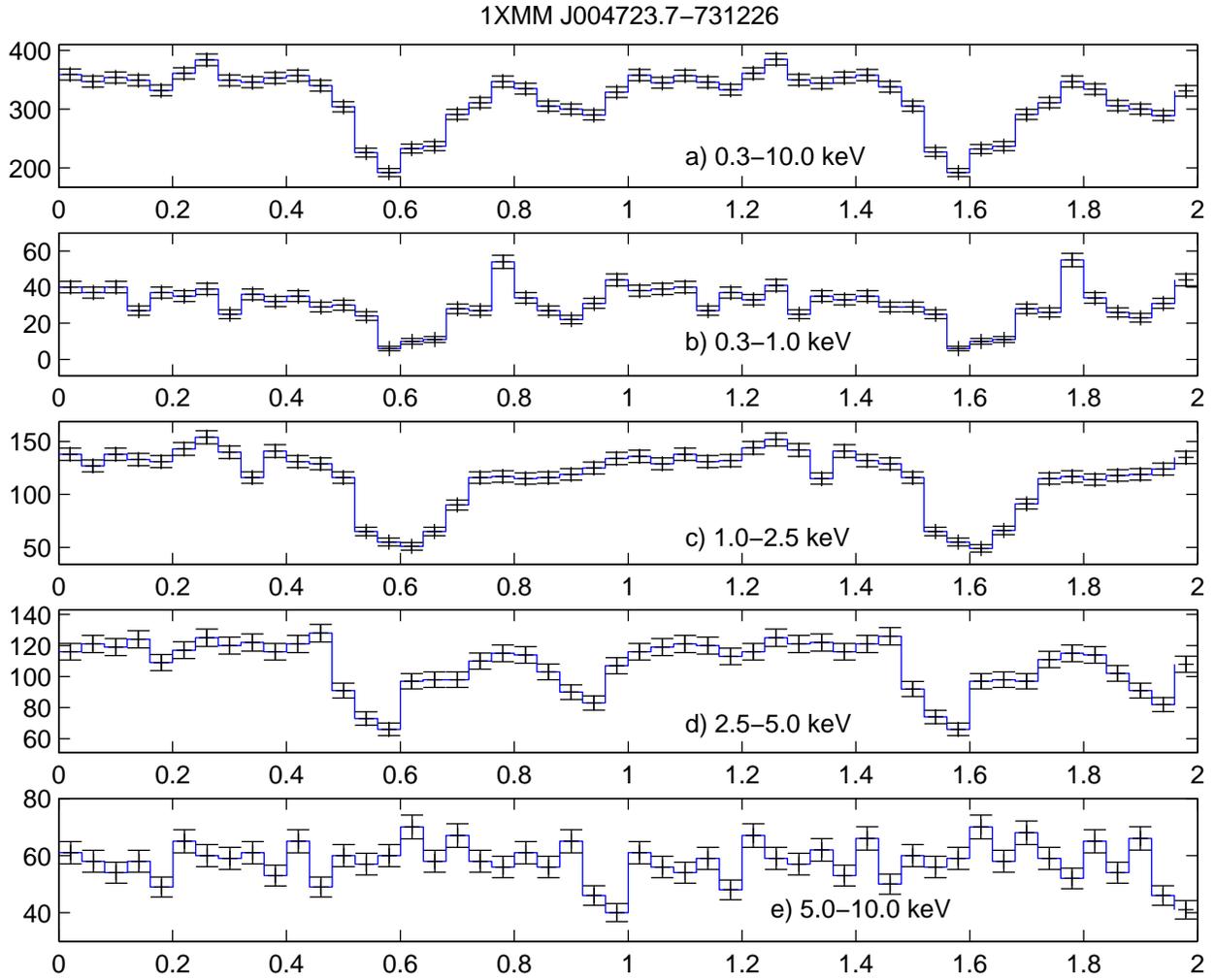}

\caption{Pulse profile for 1XMMU J004723.7-731226 for various energy
selections.  (a) 0.3-10. keV. (b) 0.3-1.0 keV. (c) 1.0-2.5 keV. (d)
2.5-5.0 keV. (e) 5.0-10.0 keV.}

\end{figure}
\subsubsection{Spectra}

\begin{figure}
\figurenum{6}
\includegraphics*[angle=270]{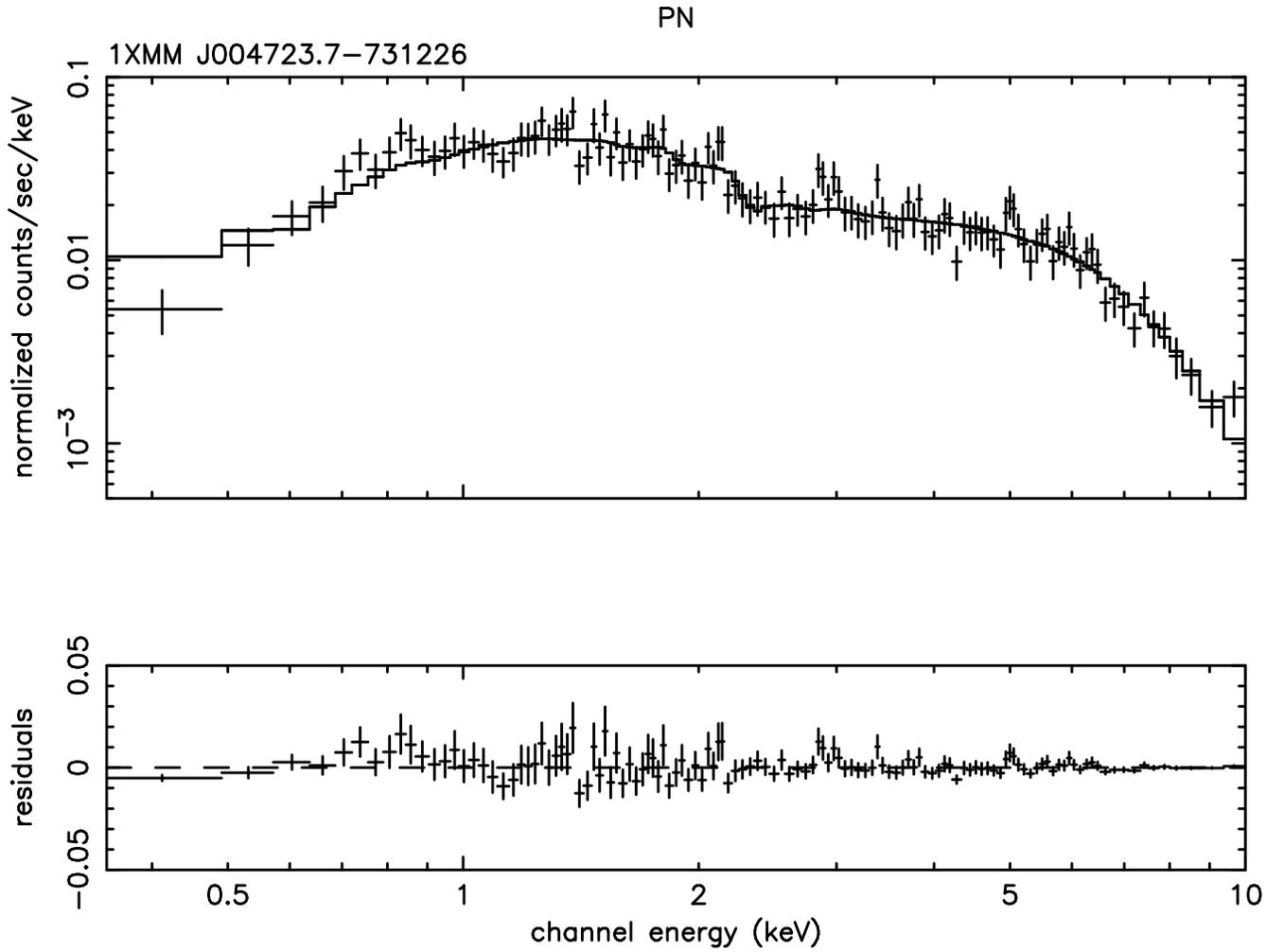}

\caption{Spectrum of 1XMMU J004723.7-731226. The fit shown is a
composite model for the PN detector consisting of powerlaw and
blackbody components, modified by neutral absorption.  Parameter
values are given in Table 2.}

\end{figure}

\begin{deluxetable}{lccccccc}
\tablecolumns{8}

\tablecaption{Spectral Parameters of Reported Pulsars.}

\tablewidth{0pt}

\tablehead{\colhead{Source} & \colhead{$\chi^{2}/(dof)$}  &
\colhead{kT(keV)} & \colhead{$N_{bb}(10^{-5}$)} & \colhead{$\alpha$} &
\colhead{$N_{pl}(10^{-4}$)} & \colhead{$n_{H}$} & \colhead{$L_{x}$(ergs/s)} }
\startdata

J004723.7-731226 & 1.02(278)& 2.2(1) & 4.6(5)& 1.7(3) & 2.2(2) &
3.4(5) & 1.1$\times 10^{36}$ \\
&	1.09(280) & - & -	& 0.74(3)  & 1.8(2) & 2.3(2)  & 1.5$\times 10^{36}$ \\

J005921.0-722317 & 0.99(20) & - & - & 1.10(16) &0.80(36)& 0.9(8)
& 3.4$\times 10^{35}$  \\

J010043.1-721134 & 0.92(30) & - & - & 2.85(8) & 3.6(4) &
0.30(2) & 2.4$\times 10^{35}$

\enddata \tablecomments{The normalization (N) of either the powerlaw
or the blackbody spectrum have units of $photons/keV/cm^{2}/s$ at 1
keV. Units of $n_{H} \times 10^{21} cm^{-2}$.  The luminosity is for
the energy interval 0.2-6 keV and an assumed SMC distance of 60 kpc.}

\end{deluxetable}

The spectral data from the three EPIC instruments were fitted
simultaneously.  We attempted to fit the spectra to various models,
using an energy bin selection that gave a minimum of 20 photons/bin.
In each trial, all model parameters were allowed to vary freely, but
were constrained to maintain the same value for each spectra.  To
account for the slightly different count rates in each detector, an
overall normalization parameter is allowed to vary for each data set.
The spectrum is best fit with a composite model consisting of
powerlaw and blackbody components, modified by neutral absorption.
(see Figure 6).  With this model we obtain an acceptable
$\chi^{2}/\nu$ = 1.02 for 278 degrees of freedom.  Best fit parameter
values for this model are listed in Table 2.
We also find an acceptable fit ($\chi^{2}/\nu$ = 1.09 for 280 degrees of
freedom) for a simple powerlaw model, and parameters for this model
are also found in Table 2.

To search for possible spectral variation as a function of
the pulsar rotation phase, we divided the data into five phase
intervals.  The phase intervals correspond to the first pulse-on
$(\phi = [0.0-0.48])$, first pulse-off $(\phi = [0.54-0.71])$, second
short pulse-on $(\phi = [0.76-0.86])$, and the second pulse-off
interval $(\phi = [0.88-0.99])$.  We subsequently divided the first
pulse-on interval into two contiguous ones, to assure that we have
comparable statistics in each phase interval.  Doing so also served
as a consistency check of our fitting procedure, which should give
comparable results for the two intervals.

The spectrum for each phase interval was then fitted to the powerlaw
plus blackbody model obtained for the overall spectrum.  We only
allowed the overall normalizations to vary, fixing the other fit
parameters to their nominal values from the overall fit.  The results
of the fitting procedure is shown in Table 3.

\begin{deluxetable}{lccc}
\tablecolumns{4}

\tablecaption{Phase Dependent Spectral Fit Results for 1XMMU J004723.7-731226.}
\tablewidth{0pt}

\tablehead{\colhead{Phase Interval} & \colhead{$\chi^{2}$/dof (dof)} &
\colhead{Null Probability} & \colhead{Flux($ergs/cm^{2}/s$)}}

\startdata

0.00-0.28 & 0.96 (104) & 0.58 & 8.75$\times 10^{-13}$\\
0.28-0.48 & 1.02 (78)   & 0.11 & 6.54$\times 10^{-13}$\\
0.54-0.71 & 3.13 (42)   & $ <10^{-10}$ & 3.38$\times 10^{-13}$\\ 
0.76-0.86 & 1.41 (33)   & 0.06 & 3.02$\times 10^{-13}$\\ 
0.88-0.99 & 0.72 (34)   & 0.89 & 3.00$\times 10^{-13}$\\ 

\enddata
\end{deluxetable}

Our results suggest that the best fit model obtained in the
overall fit is an acceptable model for the first pulse-on and the
second dip intervals.  The results for the second dip is marginal,
given the limited statistics.  However, the spectrum during the first
pulse-off region is clearly not well described by this model.  There
is a slight deficit of events in the lower energy region and a clear
excess of events in the higher energy bins above 5 keV.

Despite the limited statistics, we attempted to model the
spectrum during the first pulse-off region.  Both a simple blackbody
model as well as a simple powerlaw model (modified by neutral
absorption in the ISM) results in acceptable fits.  From the
blackbody model, we obtained a temperature of $kT = (2.3 \pm 0.2)
keV$, with a reduced $\chi^{2}$ of 1.09 (null probability of 0.32).
The simple powerlaw model resulted in an $\alpha = 0.25 \pm 0.1$,
$n_{H} = (0.4 \pm 0.2) \times 10^{22} cm^{2}$, with a reduced
$\chi^{2}$ of 1.14 (null probability of 0.25).

\section{Observations at Other Wavelengths}

Further understanding of the 202 and 263 s pulsars requires
identifying their optical counterparts.  To this end, we have searched
for stars at other wavelengths that are within a distance from the XMM
centroid location of 1.5 arcseconds plus the position error of the
putative counterpart.  The X-ray component of this radius is slightly
beyond the typical systematic error on XMM positions of 1.3 arcseconds
as given in the XMM-Newton User's Handbook.  Table 4 lists catalog
information of the best candidate for each pulsar from two different
star catalogs. For both pulsars, the two optical counterparts listed
are consistent with being the same source.  The star catalogs used are
the OGLE SMC catalog (OGLE; Udalski et al. 1998) and the Magellanic
Clouds Photometric Survey (MCPS; Zaritsky et al. 2002.)

\begin{deluxetable}{lccccccc}
\tablecolumns{8}

\tablecaption{Proposed Optical Counterparts to the Two Newly
Discovered Pulsars} \tablewidth{0pt}

\tablewidth{0pt}

\tablehead{\colhead{Source} & \colhead{Catalog} & \colhead{Cat. Num.} &
\colhead{R.A.} & \colhead{Dec.} & \colhead{Dist.(``)} & \colhead{M$_{B}$}
&\colhead{B-V}}

\startdata

1XMM J004723.7-731226  & MCPS & 1713720 & 00 47 23.42 & -73 12 27.3 & 1.28 & 16.11(2) &\
 -0.08 \\
                       & OGLE &  114733 & 00 47 23.35 & -73 12 27.0 & 1.52 & 16.10(1) &\
 -0.04 \\

1XMM J005921.0-722317 & MCPS  & 3345630 & 00 59 21.04 & -72 23 16.7 & 0.36 & 14.96(2) &\
 -0.01 \\
                      & OGLE  &  151891 & 00 59 21.04 & -72 23 17.0 & 0.10 & 14.76(2) &\
 -0.07
\enddata

\end{deluxetable}
For the 263 second pulsar, there are stars in both the OGLE and MCPS
catalogs which are 1.28 and 1.52 arcseconds away from the putative XMM
position.  The OGLE survey has a typical source localization of 0.15
arcseconds with a possible systematic error as large as 0.7
arcseconds.  For the MCPS catalog, the distributions of angular
offsets peak at 0.3 arcseconds and most stars are localized to within 1.0
arcseconds.  The angular distance between these two catalog stars is
only 0.4 arcseconds, consistent with being the same source.  As Table
4 shows, the measured magnitudes of both stars are consistent.  The B
magnitude (-2.8), color index of this star, (B-V = -0.08), and a
distance modulus for the SMC of 18.9 (Dolphin et al. 2001)) are
consistent with a B2 star (Allen's Astrophysical Quantities 2000).

The 202 second pulsar is similar, with stars in both the MPCS and OGLE
surveys being consistent in position with the XMM position and with
each other (0.3 arcseconds apart).  The OGLE counterpart
of J005921.0-722317 agrees with the OGLE counterpart for this source
given by Sasaki et al. (2003).  This tentative counterpart is
somewhat brighter with a B magnitude of -4.1 approximately consistent
with a B0 star.  In both cases, the B-V value is roughly that expected
given the galactic reddening (Hill et al. 1994).

\section{Discussion}

\subsection{The 202 s and 263 s X-ray Pulsars}

The probable optical counterparts to these sources are early B stars,
not identified as supergiants.  There is no evidence that these
possible counterparts are Be stars, however in view of the fact that a
large percentage of SMC pulsars are in this category (Haberl \& Sasaki
2000) we regard it as likely.  If they are Be HMXBs then from the
empirical relation observed for such objects by Corbet (1986) between
their pulse period and orbital period, we would expect them to have
orbital periods in excess of 100 days.

Spectra of HMXBs are usually modeled with a simple powerlaw with
photon indices in the range of 0.5-2.  For both the 202 and the 263 s
pulsars we find acceptable spectral fits with such a model.  For the
202~s pulsar the index is 1.1.  For the 263~s pulsar the index is
0.74.  This latter value is in agreement with the powerlaw fit value
of 0.76 given for this object by Haberl \& Pietsch (2004a)).  However a
slightly better spectral fit to the 263 s pulsar data is provided by a
powerlaw plus a black body model with a black body temperature of 2.2
keV.  See Table 2.  This temperature is significantly higher than that
found for other SMC HMXBs modeled with an additional blackbody term.
The phase-resolved spectrum of this source reveals a hard component
during the first pulse-off interval (phase 0.54-0.71) which can be
modeled with either a simple powerlaw (index 0.25) or a blackbody
model (temperature 2.3 keV).

\subsection{Possible AXP--CXOU J010043.1-721134}

If the possible AXP is located in the SMC at a distance of 60 kpc, its
luminosity is $2.4 \times 10 ^{35}$ ergs/s.  This luminosity is well
within the range of luminosities given for AXPs (see for example
Mereghetti et al. 2002).  This luminosity is $\sim$ 50\% larger than
the average luminosity given for this source in the compilation of
luminosities given by Lamb et al. 2002b.  Given that this luminosity
is obtained with a different instrument than those of given in Lamb et
al., we do not regard this feature as disqualifying the candidacy of
the source as an AXP which generally have near constant luminosity.
The observed spin-down time scale of the source, $P/\dot P$, of $\sim$
10 kyr is similar to the values obtained for other AXPs which are in
the range of $10^{4-5.5}$ yr.  By establishing the correct period to
be 8.02 s rather than 5.4 s we remove what would have been the shortest
period AXP and replace it with a candidate AXP whose period is nearly
equal to the average of the 6 AXPs given in Mereghetti et al.(2002).

\section{The High Mass X-ray Binary Pulsars of the SMC}

In Table 5 we list all of the HMXB pulsars of the SMC which have been
optically identified.  The list is compiled from the HMXB list of Liu
et al. (2000) with additional pulsars from ASCA observations
summarized by Yokagawa et al. (2003), an additional pulsar from
XMM-Newton observations by Sasaki et al. (2003) plus the two newly
discovered pulsars discussed in this paper.  In addition, in Table 6
we list all other X-ray pulsars of the SMC of which we are aware.
Some of these may in fact by HMXBs but they lack an optical
identification.

In Figure 7 we have plotted the locations of the pulsars of Table 5
and 6 overlaid on contours of recent star formation in the SMC taken
from Figure 7 of Maragoudaki et al. (2001).  The contours show the
density of young stars having ages $(1.2-3)\times 10^7$ yr.  (Note:
Our Figure 7 is similar to Figure 12 of Yokogawa et al. (2003),
however, we have chosen to plot only pulsars confirmed through the
detection of pulsations rather than including candidate HMXBs as
Yokogawa et al. do.)  In view of the formation history of HMXBs
(Verbunt \& van den Heuvel 1995) the correspondence between the young
stars and HMXBs is to be expected.  If there were significant
discrepancies between the spatial locations of the HMXBs and regions
of young stars, then our understanding of the evolution of X-ray
binaries would need modification.  A conclusion to be drawn from this
figure is that most of the X-ray pulsars of Table 6 not yet identified
as HMXB pulsars may be high mass binaries.

\begin{deluxetable}{lcccc}

\tablecolumns{5}

\tablecaption{High Mass X-ray Binaries in the SMC}

\tablewidth{0pt}

\tablehead{\colhead{Source} & \colhead{Period (s)} & \colhead{Type}
&\colhead{Flux Density\tablenotemark{x}} & \colhead{References}}

\startdata

   0115-737 (SMC X-1) & 0.71  & & &  a \\
        SMC X-2 &       2.37  & Be & 7.6  & b \\
   J0059.2-7138 &       2.76  & Be & 18   &   a,b     \\
   J0051.8-7231 &       8.90  & Be & 0.34   &    a,b   \\
   J0052.1-7319 &       15.3  & Be  & 12.3 & a,b   \\
   J0117.6-7330 &       22.1  & Be & 9.2   &    a,b   \\
   J0111.2-7317 &       31.0  & Be & 16   &   a,b   \\
   J0053.8-7226 &       46.6  & Be & 0.15  &  a,b    \\
RX J0054.9-7226 &       59.0  & Be & 0.76  & b  \\
      J0049-729 &       74.7  & Be & 0.49  &  a,b   \\
      J0051-722 &       91.1  & Be& 3.0   &  a,b   \\
XMMU J005605.2-722200 & 140.  & & & c \\
CXOU J005750.3-720756 & 152.  & & & d \\
      J0054-720 &       169.  & Be & 4.3  &  a,b   \\
AX J0051.6-7311 &       172.  & Be & 0.19  & b \\
1XMMU J005921.0-722317 &        202.  & & & e \\
1XMMU J004723.7-731226 &        263.  & & & e \\
   J0058-7203   &       280. & &   &  a,b   \\
CXOU J010102.7-720658 & 304. & &  & d \\
   J0050.7-7316 &       323. & Be & 0.19   &  a,b  \\
1SAX J0103.2-7209 &     349. & Be & 0.14   & b \\
RX J0101.3-7211 &       455. & Be & 0.28   & b \\
CXOU J005736.2-721934 & 565. & & & d \\
AX J0049.5-7323 &       756. & & & b

\enddata
\tablenotetext{x}{Maximum X-ray flux density in units of $\mu$Jy.
1 $\mu$Jy = 2.4$\times10^{-12}$ ergs/s/cm$^2$/keV.}

 \tablecomments{The maximum X-ray flux density is calculated from flux
values cited in the original papers given in the references.  We
approximate the flux density as flux divided by the energy interval
over which the flux is measured, generally 0.1 to 2.5 keV (ROSAT) or
2.0 to 10. keV (ASCA and Chandra).  This approximation is valid if the
source spectrum has a powerlaw dependence with number index -1.
References: a=Liu et al. (2000), b=Yokagawa et al. (2003), c=Sasaki et
al. (2003), d=Macomb et al. (2003), e=This work}

\end{deluxetable}
\begin{figure}
\figurenum{7} \epsscale{0.65} \plotone{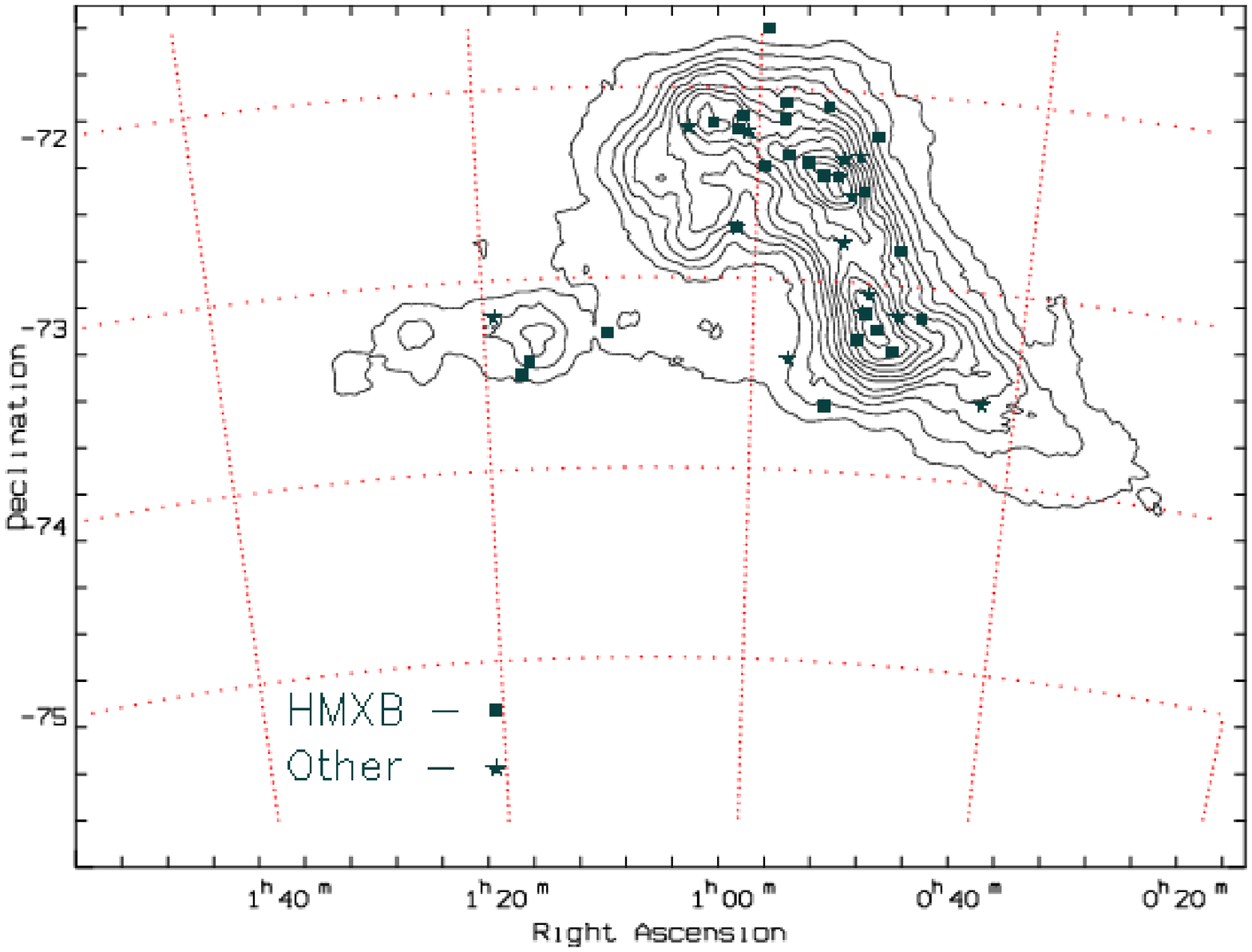}

\caption{Locations of the HMXBs of the SMC relative to its young star
forming regions.  The contours are those of Figure 7 of Maragoudaki et
al.  (2001), corresponding to contours of constant star density for
stars with ages $(1.2-3)\times10^7$ yr.  Note that XTE positions (nine
of the stars) are uncertain by as much as a degree.}

\end{figure}

\begin{deluxetable}{lcr}

\tablecolumns{3}

\tablecaption{Other X-ray Pulsars in the SMC}
\tablewidth{0pt}

\tablehead{\colhead{Source} & \colhead{Period (s)} &
 \colhead{References}}

\startdata

AX J0043-737    &       0.0876  &       Yokagawa \& Koyama (2000)\\
XTE J0119-731   &       2.17    &       Corbet et al. (2003a) \\
AX J0105-722    &       3.34    &       Yokogawa \& Koyama (1998)\\
XTE J0052-723   &       4.78    &       Corbet et al. (2001)\\
XTE J0103-728   &       6.85    &       Corbet et al. (2003b) \\
XTE Position A  &       7.77    &       Corbet et al. (2003c)\\
CXOU J010043.1-721134 &   8.02  &       Lamb et al. (2002b, 2003)\\
AX J0049-732    &       9.13    &       Imanishi et al. (1998)\\
XTE J0050-732\#1 &       16.6    &       Lamb et al. (2002a)\\
XTE J0055-727    &	18.37	&	Corbet et al. (2003d)\\
XTE J0050-732\#2 &       25.5    &       Lamb et al. (2002a)  \\
CXOU J005527.9-721058 &	34.08	&	Edge et al. (2004a)\\
XTE Position A and D & 46.4 	& 	Corbet et al. (2004)\\
XTE J0052-725 	&	82.4	 & 	Corbet et al. (2002)\\
XTE Position A 	& 	89 	& 	Corbet et al. (2004)\\
XTE SMC95 	& 	95.0 	& 	Laycock et al. (2002)\\
AX J0057.4-7325 & 	101. 	 & 	Yokogawa et al. (2000)\\
CXOU J0053238-722715 &	138.	&	Edge et al. (2004b)\\
XTE J0103-728	 & 	144.	& 	Corbet et al. (2003b)\\
XTE Position B 	& 	164.7	 & 	Corbet et al. (2004)\\
CXOU J005455.6-721510 & 499/503) &      Edge et al. (2004b), Haberl et al. (2004)\\
XMMU J005517.9-723853 &	701.6	&	Haberl et al. (2004)

\enddata

\end{deluxetable}

The overabundance of the SMC HMXB pulsar population relative to that
of the Milky Way has been noted many times before (cf Yokogawa et
al. 2003).  For example in the compilation of HMXBs of Liu et
al.(2000) there are 50 HMXB pulsars listed for the Milky Way whereas
for the SMC with approximately 1\% of the mass of the Milky Way we
have compiled a list of 24 HMXB pulsars (Table 5).  There are
absorption and sky coverage effects that may diminish the number of
detected HMXBs for the Milky Way, nevertheless, on the face of it the
SMC population of HMXBs is overabundant relative to the Milky Way by a
factor of $\sim 50$.  This points to a spectacular episode of star
formation in the SMC beginning approximately 30 million years ago.

One of the advantages of population studies with the SMC is its well
known distance.  Thus fluxes are proportional to luminosity, a
fundamental parameter of each binary system.  We have investigated the
possibility of a correlation between the spin period of the neutron
star and its luminosity using the maximum reported flux density.  We
have restricted this investigation to the subset of the HMXBs in Table
5 which are firmly identified with Be companions.  On the basis of a
simple model of the accretion process for such objects, discussed
below, one might expect that rapid rotation leads to greater maximum
luminosity.  We find empirically that notion is supported by the
observations.

In Figure 8(a) we have plotted the maximum X-ray flux density versus
spin period for the Be HMXBs.  (Note: we have added four Be HMXB
pulsars from the LMC and multiplied their flux densities by 0.727 to
account for the difference between the distance to the LMC and to the
SMC.  The LMC sources are: 0535-668, J0544.1-7100, J0502.9-6626 and
J0529.8-6556.)  The maximum flux densities are anti-correlated with
spin periods, having a correlation coefficient of -0.75.  The
probability that there is no correlation between spin period and
maximum flux density is $2\times10^{-4}$.  The solid line in Figure
8(a), a least square fit to the points, has a slope of -0.73$\pm0.16$.
A similar anti-correlation for all X-ray binary pulsars was noted by
Stella, White \& Rosner (1986) and interpreted as the result of
centrifugal inhibition of accretion.  We note that if the 202 and 263
s pulsars are also Be HMXBs, and if we take the fluxes determined in
this observation as maximum (certainly a lower limit), they may be
located as well on the plot of Figure 8 (a).  Doing this, we find that
the point for the 263 s pulsar lies below the best fit straight line
of Figure 8 (a) by a factor of two; the 202 s pulsar lies somewhat
lower by about a factor of 6.  In view of the scatter of points shown
in Figure 8 (a) these differences are relatively minor.

We find that a similar correlation exists for the Be HMXB pulsars of
the Milky Way.  Haberl \& Sasaki (2000) list the spin period and
maximum luminosity of 27 HMXB pulsars of the Milky Way (their Table
3).  We have plotted these data in Figure 8(b).  The data are strongly
correlated, correlation coefficient of -0.69, with a probability of
$6\times10^{-5}$ that the correlation is by chance, and a fitted slope
of -1.01$\pm0.21$.  We have combined the data for the Milky Way and
for the Magellanic clouds into a single plot, shown in Figure 9.  We
have scaled the Magellanic cloud data to an SMC distance of 60 kpc.
The correlation coefficient for this combined data, Milky Way plus the
Magellanic Clouds (46 points), is -0.71 with a probability that it is
due to chance of 2.9$\times10^{-8}$.  The fitted slope has a value of
-0.95$\pm0.14$.

Many HMXBs are transients with large variation in flux.  This fact
undoubtedly contributes to the scatter of the points of Figure 8 and
9.  Nevertheless there is a significant inverse correlation between
maximum X-ray flux density and period.  This correlation should be
understandable in terms of the physics of the accretion process.  In a
simple model of accretion from the stellar wind of the Be companion,
the rate of accretion onto the neutron star is proportional to the
density of the wind, which depends on 1/r$^n$ where r is the distance
between the companion and the neutron star. ``n'' is found to range
from 2 to 4, but mostly between 2.5 and 3 (Waters \& van Kerkwijk
1989).  We take n=3.  By Kepler's law, r goes as P$_{orb}^{2/3}$ of
the neutron star.  Empirically spin period for Be HMXBs is
proportional to P$_{orb}^{2}$ (Corbet 1986).  Thus luminosity should
scale roughly as P$_{spin}^{-1}$, implying a slope of -1 for the
correlation of the data points of Figures 8 and 9 in rough agreement
with what is observed.  Note that only four of the Be systems listed
in Table 5 have measured orbital periods.  P$_{spin}$, because of the
Corbet relation, thus becomes a surrogate for P$_{orb}$.

Of course there are many complications to this very simple analysis.
One complication is that, for highly elliptic orbits, mass transfer
from the companion to the neutron star will occur near periastron and
thus the appropriate scaling distance is significantly less than r.
Nevertheless, further work on this demonstrable correlation between
spin period and X-ray luminosity may be fruitful.  It could be that a
suitably improved relation would serve as a distance indicator.  We
have found that the maximum degree of correlation between the
Magellanic Clouds and the Milky Way is achieved if the luminosities
are reduced by roughly 2/3 of their value, which would imply a
distance reduced by $\sim$18\%.  Since we believe that the distances to
the LMC and the SMC are rather well determined a more likely
possibility is that the Magellanic Be HMXBs are perhaps more luminous
that those of the Milky Way.  Another possibility is that the
estimates of the distances of the Milky Way sources are inaccurate,
and their luminosities in general need to be increased.

\begin{figure}
\figurenum{8} \epsscale{1.0} \plotone{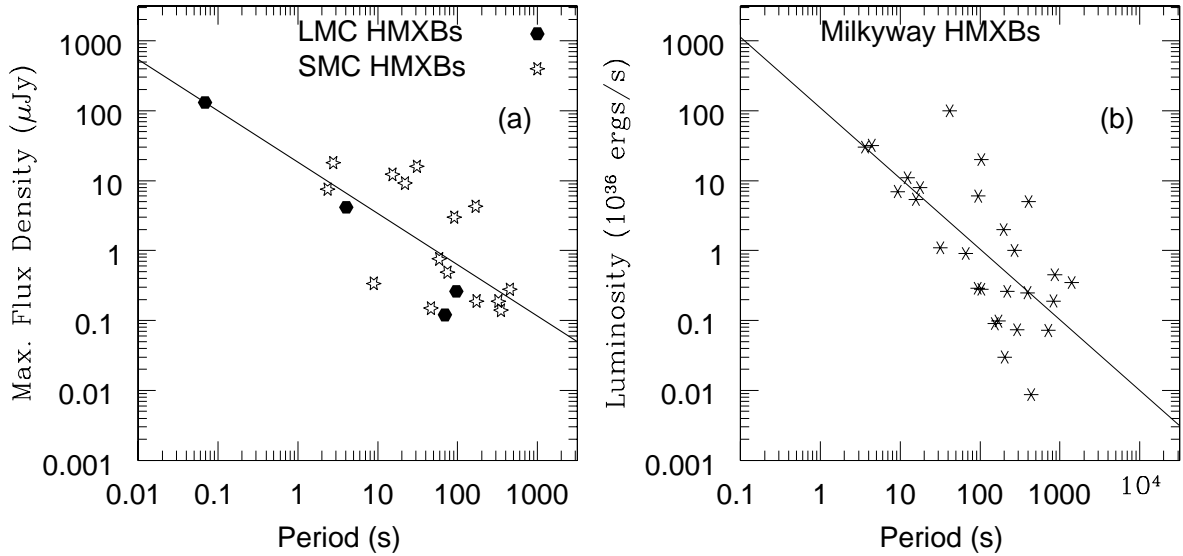}

\caption{ a)Maximum flux density versus pulse period for the Be HMXB
pulsars of the SMC and LMC.  The flux densities of the 4 LMC Be HMXBs
have been scaled to the distance of the SMC.  The solid line is a
least square fit to the 19 data points with a slope of -0.73$\pm0.16$.  The
correlation coefficient of these data is -0.75 with a probability that
it occurs by chance of $2\times10^{-4}$. b) The Milky Way Be HMXB
pulsars from Haberl \& Sasaki (2000).  The solid line is a fit to the
27 data points with a slope of -1.01$\pm0.21$.  The correlation coefficient
for these data is -0.69 with a probability that it occurs by chance of
$6\times10^{-5}$.}
\end{figure}

\begin{figure}
\figurenum{9} \epsscale{1.0} \plotone{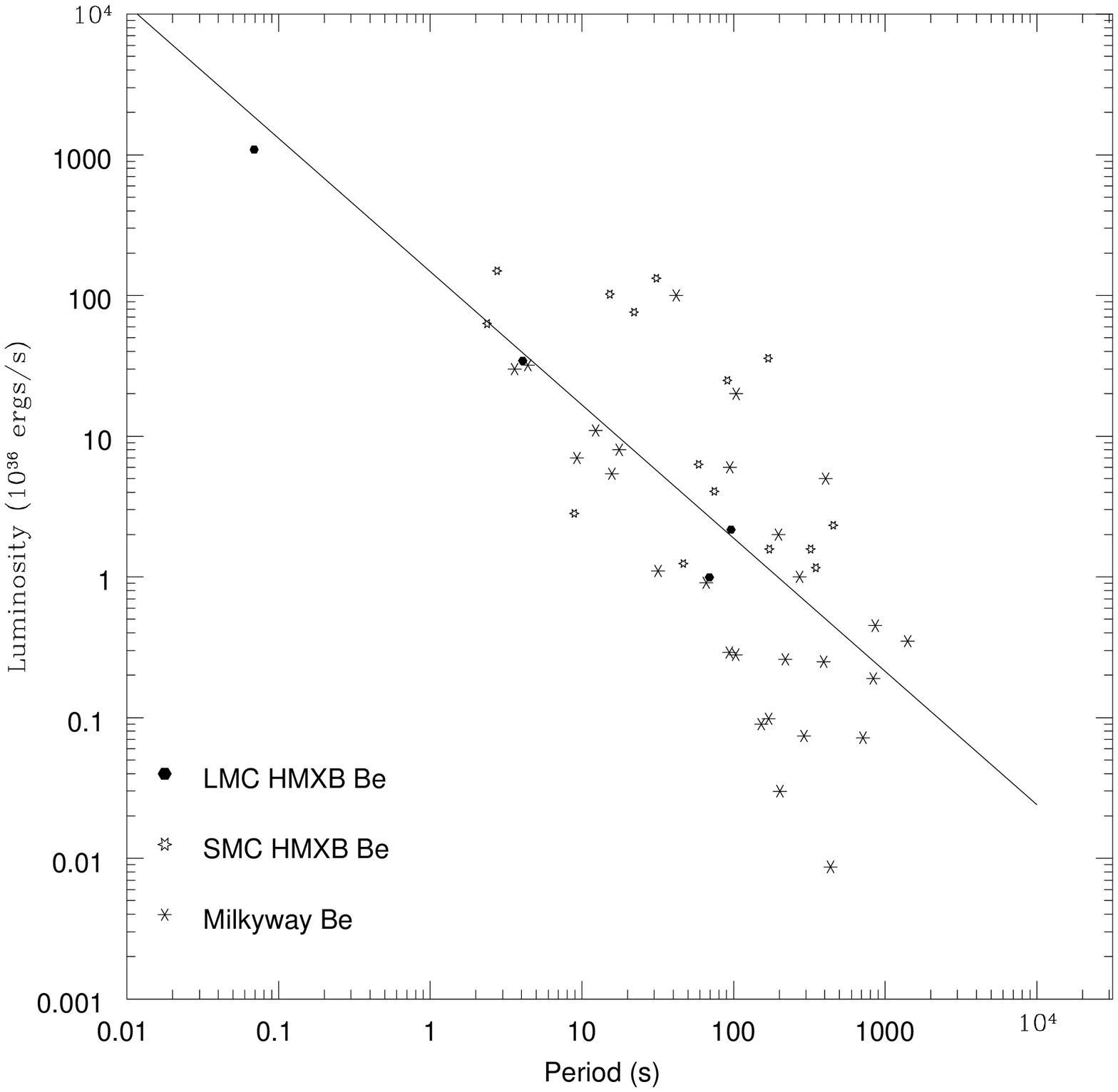}

\caption{Maximum X-ray luminosity for 46 Be HMXB pulsars of the Milky
Way, LMC and SMC.  For the Magellanic Cloud pulsars the luminosity is
for the interval 2-10 keV, an interval believed comparable the
intervals for the data of the Milky Way (Haberl \& Sasaki 2000).  The
pulsar data of the Magellanic Clouds has been scaled to an SMC distance
of 60 kpc.  The correlation has a probability of occurring by chance of
$2.9\times10^{-8}$.  The line, a fit to the data points, has a
slope -0.95$\pm0.14$.}

\end{figure}

We have compared the period distribution of SMC population of HMXB
pulsars to that of the Milky Way.  In Figure 10, the cumulative
distributions of pulse periods are shown for the HMXBs for both
galaxies.  For the Milky Way the distribution is compiled from the
catalog of Liu et al. (2000).  For the SMC the distribution is
compiled from Table 5.  The distributions are normalized to the total
number of HMXBs of each kind.  The two distributions are virtually
identical for periods less than 200 s, however for longer periods
there is a slight deficit of SMC HMXBs relative to the Milky Way.  If
the SMC population is indeed younger than the Milky Way population, a
preference for short-period pulsars in the SMC could be construed as
evidence that high-mass binaries spin-down over time on average.
Since our data does not currently support such a difference, the
interpretation is that significant changes of pulse period
distributions occur only over timescales longer than the difference in
average age between the SMC and Milky Way populations.  (This
discussion will need to be amended if a significant number of the
pulsars in Table 6 should turn out to be HMXBs, since their periods
are predominately short period.  For example, 12 of the 14 pulsars in
Table 6 have periods less than 100s, whereas only 11 of 24 of the
pulsars in Table 5 have periods less than 100s.)  The slight deficit
of long period SMC pulsars seen in Figure 10 may be due to their
relative faintness compared to their more rapidly spinning cousins
(see Figure 8(a)).  This argues for even deeper surveys of the SMC to
establish a complete catalog of such objects.

\begin{figure}
\figurenum{10} \epsscale{0.65} \plotone{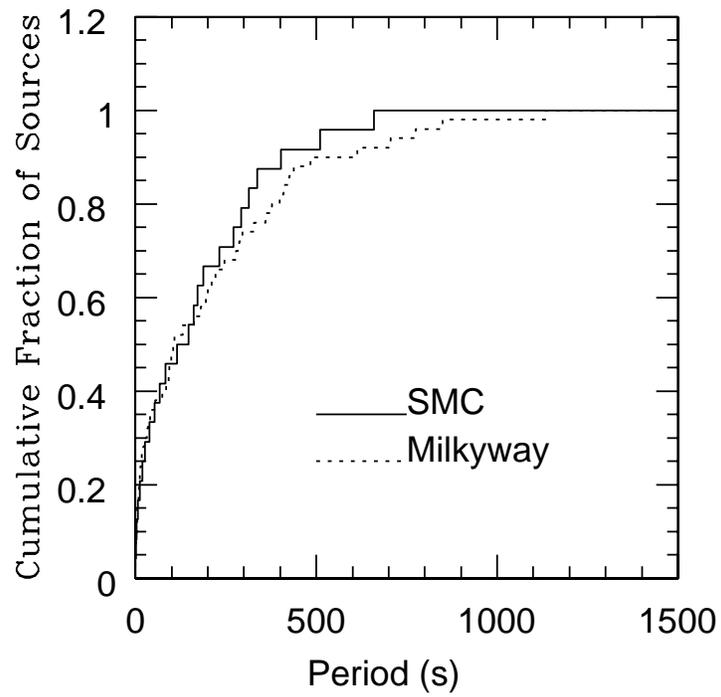}
\caption{Pulse period distributions for the Milky Way and Small
Magellanic Cloud high-mass X-ray binary populations.  The two distributions
are virtually identical for periods less than 200 s, however for longer periods
there is a slight deficit of SMC HMXBs relative to the Milky Way.}

\end{figure}

\acknowledgements

Helpful comments on the manuscript and insight into the accretion
process have been made by T. A. Prince.  This work made use of
software and data provided by the High-Energy Astrophysics Archival
Research Center (HEASARC) located at Goddard Space Flight Center, as
well as SAS data analysis software from the the XMM-Newton team.
Additional use was made of the SIMBAD catalogue at CDS, Strasbourg
France, and NASA's Astrophysics Data System (ADS).  This research was
carried out at Caltech, Boise State University and the Jet Propulsion
Laboratory and supported by the Jet Propulsion Laboratory, California
Institute of Technology, under a contract with the National
Aeronautics and Space Administration. .

\end{document}